\documentclass{article}
\newcommand{\dddot}[1]{\stackrel{\mathbf{\ldots }}{#1}}

\begin{document}

\title{Comments on Bouda and Djama's ``Quantum Newton's law"} 

\author{Edward R. Floyd \\
10 Jamaica Village Road, Coronado, CA 92118, USA\\
floyd@mailaps.org}

\maketitle

\abstract{Discussion of the differences between the trajectory representation of Floyd and 
that of Bouda and Djama [Phys.\ Lett.\ {\bf A 285} (2001) 27] renders insight: while 
Floyd's  trajectories are related to group velocities, Bouda and Djama's are not.  Bouda 
and Djama's reasons for these differences are also addressed.}

\bigskip



\clearpage

Bouda and Djama have recently presented trajectories for non-relativistic quantum 
mechanics and have noted that their trajectories differ with mine.\  [\ref{bib:bd}]\ \   
Bouda and Djama have offered two explanations why these trajectories differ.\ 
[\ref{bib:bd}]\ \  Herein, I comment on the findings of Bouda and Djama, present insight 
into the differences between the trajectory representations, and offer a different 
explanation why these trajectories differ.

The insight between the trajectory representations is manifested in Bouda and Djama's  
application to the free particle in one dimension in Ref.\ \ref{bib:bd}.  Let us now 
examine their application.  Their equation of motion for the free particle is their Eq.\ (34), 
which they cite as ``quantum time equation for the free particle."\ [\ref{bib:bd}]\ \   
Equation (34) may be manipulated so that 

\setcounter{equation}{33}
\renewcommand{\theequation}{B\&D's\ \arabic{equation}}

\begin{eqnarray}
\frac{\hbar}{(2mE)^{1/2}} \arctan [a \tan (2Et/\hbar) +  b] & = & x(t) - x_0 \\ 
a \tan (2Et/\hbar ) + b & = &  \tan \left[ \frac{(2mE)^{1/2}}{\hbar }(x(t)-x_0) \right] 
\nonumber \\ 
2Et & = & \underbrace{\hbar \arctan \left[ \frac{1}{a} \tan \left( 
\frac{(2mE)^{1/2}}{\hbar}(x(t)-x_0)\right) - \frac{b}{a}\right]}_{\displaystyle S_0} 
\nonumber 
\end{eqnarray}

\noindent where an intermediate step has been included for clarity.   We use the notation 
of Bouda and Djama:  $(E,a,b,x_0)$ are constants of the motion where $E$ is also 
energy; $S_0$ is the quantum reduced action.  What Bouda and Djama have done in 
Ref.\ \ref{bib:bd} is to turn the quantum reduced action inside out.  By Eq.\ (B\&D's 34), 
$S_0 = 2Et$.  Then, Bouda and Djama's Eq.\ (B\&D's 34) is not only their equation of 
motion but also the quantum reduced action, $S_0$, a generator of the motion.  Bouda 
and Djama have presented $S_0$, for discussing my trajectories, by their Eq.\ (35) as

\begin{equation}
S_0 = \hbar \arctan \left[ a_{35}\tan \left( \frac{(2mE)^{1/2}x}{\hbar }\right) + 
b_{35}\right] + \hbar \lambda
\end{equation}

\setcounter{equation}{0}
\renewcommand{\theequation}{\roman{equation}}

\noindent where for explicitness the integration constants $a$ and $b$ for their Eq.\ 
(B\&D's 35) have been subscripted by 35.  For consistency between Bouda and Djama's 
Eqs.\ (B\&D's 34) and (B\&D's 35), we must set $a_{35}=1/a$ and $b_{35}=-b/a$.

On the other hand, Bouda and Djama have presented the equation of motion for my 
corresponding trajectory for the free particle by their Eq.\ (36).  Their Eq.\ (36) may be 
presented in terms of $(a,b)$ rather than $(a_{35},b_{35})$ as

\begin{equation}
t-t_0 = a \frac{(2m/E)^{1/2} x}{(a^2 + b^2 +1) + \sigma \cos[2(2mE)^{1/2} x/\hbar - 
\gamma]}
\label{eq:eom}
\end{equation}

\noindent where

\[
\sigma = (a^4+b^4+1+2a^2b^2 +2b^2-2a^2)^{1/2},\ \ \ \gamma = \arctan \left( \frac{-
2b}{a^2+b^2-1} \right).
\]

The velocity of the particle that has been described by the equation of motion, Eq.\ 
(\ref{eq:eom}), is given by $\dot{x} = dx/dt = (dt/dx)^{-1} = (\partial t/\partial x)^{-1}$ 
as the right side of Eq.\ (\ref{eq:eom}) does not contain $t$ explicitly.  The particle 
velocity can be expressed  by  Eq. (\ref{eq:eom}) as  $\dot{x}= (\partial ^2 S_0/\partial E 
\partial x)^{-1}$. We can change the order of differentiation between $x$ and $E$.  
Thus, the particle velocity is given by $\dot{x} = \big[ \partial (\partial S_0/\partial 
x)\big/ \partial E\big] ^{-1}$.\ [\ref{bib:fm}--\ref{bib:c1}] \ \  For the case at hand, the 
velocity for the free particle is given by

\begin{equation}
\dot{x} = \frac{\partial E}{\partial (\partial S_0/\partial x)} = \frac{1}{a} \frac{{\cal 
A}^2}{ (2m/E)^{1/2} {\cal A} + 2m \sigma \sin \{[(2mE)^{1/2}(x-x_0)/\hbar] - \gamma 
\}(x-x_0)/\hbar }
\label{eq:v}
\end{equation}

\noindent where ${\cal A} = 1 + a^2 + b^2 + \sigma \cos \{[(2mE)^{1/2}(x-x_0)/\hbar] - 
\gamma \}$.

Let us examine Eq.\ (\ref{eq:v}).  The form $\partial E \big/ \partial (\partial S_0/\partial 
x)$ is reminiscent of the canonical equation of Hamilton for $\dot{x}$ as given by Bouda 
and Djama's Eq.\ (12).  We recall that Hamilton's principal function, $S$, propagates 
like a wave in configuration space [\ref{bib:g}] with angular frequency $E$ and wave 
number $S_0$.  We see that $\dot{x}$ as given by Eq.\ (\ref{eq:v}) is a group velocity 
because it is rendered by the partial derivative of angular frequency, $E$, with respect to 
the wave number or conjugate momentum, $\partial S_0/\partial x$.  (Lest we forget, the 
conjugate momentum in not the mechanical momentum, i.e., $\partial S_0/\partial x \ne 
m\dot{x}$.)  The group velocity describes the particle's velocity as well as the 
propagation of the envelop of the $S$-wave in configuration space.  We note that, for a 
given energy $E$, the group velocity is dependent on the particular trajectory or 
microstate specified by $(a,b)$.  The coefficients $(a,b)$ are determined by the initial 
conditions for the quantum stationary Hamilton-Jacobi equation (QSHJE), Bouda and 
Djama's Eq.\ (3), and for the particular set of independent solutions $(\phi _1,\phi _2)$ of 
the stationary Schr\"odinger equation chosen by Bouda and Djama.  On the other hand, 
Bouda and Djama's particle velocity, $\dot{x}_{bd}$ is given by their Eq.\ (17) as

\setcounter{equation}{16}
\renewcommand{\theequation}{B\&D's\ \arabic{equation}}
 
\begin{equation}
\frac{\partial S_0}{\partial x} = \frac{2(E-V)}{\dot{x}_{bd}}. 
\end{equation}

\setcounter{equation}{2}
\renewcommand{\theequation}{\roman{equation}}

\noindent For the free particle in terms of $(a,b)$ and not $(a_{35},b_{35})$, 
$\dot{x}_{bd}$ is given by  

\[
\dot{x}_{bd} = \frac{2E}{\partial S_0/\partial x} = \frac{1}{a} \left( 
\frac{E}{2m}\right) ^{1/2} {\cal A}, 
\]

\noindent  which is neither a group nor a phase velocity.  Herein, phase or wave velocity 
describes the propagation of the wave fronts of constant $S$ in configuration space.\ 
[\ref{bib:g}]\ \  For $a=1$ and $b=0$, $\dot{x}$ and $\dot{x}_{bd}$ both reduce to the 
classical particle velocity $(2E/m)^{1/2}$ for the free particle.  

Nonlocality is also manifested by Eq.\ (\ref{eq:v}).  For $a \ne 1$ or $b \ne 0$, as the 
value $(x-x_0)$ becomes large, $\dot{x}$ becomes infinite at certain locations.  This and 
other related implications of nonlocality in Eq.\ (\ref{eq:v}) have been previously 
reported.\ [\ref{bib:f94}] \ \ Nevertheless, $\dot{x}$ remains integrable as manifested by 
Eq.\ (\ref{eq:eom}).  This nonlocal effect is not as great as the nonlocal effect that exists 
in the classically forbidden region of bound states.\ [\ref{bib:f00a}] \ \ For the free 
particle, the measure of where $\dot{x} = \infty$ is sufficiently small so that the particle 
transverses only a finite distance over any duration that includes when $\dot{x} = \infty$.  
For bound states the particle transverses an infinite distance in a finite duration when it 
transverses from the classical WKB turning point out to $x=\pm \infty$ where it reverses 
and returns to the classical WKB turning point.\ [\ref{bib:f00a},\ref{bib:f82}]

Bouda and Djama offer two arguments why the two trajectory representations should 
differ.  In their first argument, Bouda and Djama reported that in my approach, it is only 
in the classical limit, $\hbar \to 0$, that $S_0 = 2E(t-t_0)$ from which $S$ manifests the 
integration of the Lagrangian over time.  This is a misunderstanding.  For clarity, I only 
showed in my examination of the classical limit in Ref.\ \ref{bib:f00} that in the classical 
limit $\hbar \to 0$ one has $S_0=2E(t-t_0)$.   I did not examine whether $S$ was the 
integral of the Lagrangian for $\hbar \neq 0$.  Nevertheless, my findings can be 
generalized.  The quantum Lagrangian, $L$, is defined to be the time derivative of the 
quantum Hamilton's principal function, which is a generator of motion in $(x,t)$ space.  
The quantum Lagrangian may be formally constructed, in principle, from the quantum 
Hamilton's principal function by 

\begin{eqnarray*}
\frac{dS}{dt}& = &\frac{\partial S}{\partial x}\dot{x} + \frac{\partial S}{\partial t}\\
& = & \frac{\partial S}{\partial x}\dot{x} - \frac{1}{2m}\left( \frac{\partial 
S}{\partial x}\right) ^2 - V(x) - \frac{\hbar^2}{4m} \left[ \frac{\partial ^3S/\partial 
x^3}{\partial S/\partial x} - \frac{3}{2}\left( \frac{\partial ^2S/\partial x^2}{\partial 
S/\partial x}\right) ^2 \right] \equiv L.
\end{eqnarray*}
 
\noindent The quantum Lagrangian has not yet been brought into the form 
$L(\dddot{x},\ddot{x},\dot{x},x)$.  We suspend development of a Lagrangian 
representation of quantum motion here.  I do not recommend reverting to a Lagrangian 
representation to resolve quantum motion.  A Lagrangian representation implies using a 
variational principle to minimize transit time for the propagation of  S in accordance with 
Fermat's principle.  As such, it does not render group motion.  Instead, it renders phase 
motion developed from phase velocities, accelerations and jerks.  We shall give more on 
this later.  Furthermore, as noted by Bertoldi, Faraggi and Matone for time independence, 
the quantum equivalence principle (QEP) is implemented in a Hamilton-Jacobi 
representation rather then at the level of equations of motions.\ [\ref{bib:bfm}]  

Since $S_0$ is explicitly the quantum reduced action and not the classical reduced action,  
Bouda and Djama, in their second argument, express some reservation about the validity 
of using Jacobi's theorem with $S_0$ to furnish the constant time coordinate, $t_0$, 
while rendering time parameterization.  Carroll previously put their reservation to rest 
elsewhere for he had shown that Jacobi's theorem is applicable to the quantum reduced 
action as the quantum reduced action is a Legendre transformation of the negative of the 
quantum Hamilton's principal function.\ [\ref{bib:c},\ref{bib:c1}] \ \ Let us now 
illustrate.  Following Carroll, we consider a stationary state where $t$ does not explicitly 
appear in the quantum Hamiltonian.  We may then separate Hamilton's principal 
function, $S$, so that $S=S_0-Et$.  Then $\partial S/\partial t = -E$ so that $t=t(E)$.  We 
let ${\cal S} = -S$ where $\partial {\cal S}/\partial t =E$.  It follows that $S_0 = Et-{\cal 
S}$ or 

\begin{equation}
S_0 = t\frac{\partial {\cal S}}{\partial t} - {\cal S}
\label{eq:L1}
\end{equation}

\noindent where $S_0$ is the Lagrangian transformation of ${\cal S}$.   With $t$ the 
active variable and $x$ the passive variable as expressed by Lanczos [\ref{bib:lanczos}], 
we hold the passive $x$ variable fixed to give $\partial S_0/\partial E = t + E (\partial 
t/\partial E) - (\partial {\cal S}/\partial t)(\partial t/\partial E)  = t$.  It follows that

\begin{equation}
{\cal S} = E\frac{\partial S_0}{\partial E} - S_0,
\label{eq:L2}
\end{equation}

\noindent where ${\cal S}$ is the Lagrangian transform of $S_0$.  Equations 
(\ref{eq:L1}) and (\ref{eq:L2}) form the Lagrangian dual transformations between ${\cal 
S}$ and $S_0$.  The quantum reduced action, $S_0$, is a function of $E$ instead of $t$, 
so Jacobi's theorem is applicable to render time parameterization.

I now offer two explanations why the two trajectory representations differ.  First, for my 
trajectory representation, the propagation of Hamilton's principal function, $S$, in 
configuration space is determined by Fermat's principle where the time for the surface of 
constant $S$ to transit between two points is an extremum.  As discussed by Park 
[\ref{bib:park}], the transit  time for phase wave, and not the signal, is stationary (usually 
minimized).  The phase wave propagates with the phase velocity while the signal (the 
wave envelope) propagates with group velocity, which is developed from Jacobi's 
theorem.  The fundamental reason that the equations of motion, Eqs.\ (\ref{eq:eom}) and 
(B\&D's 17), differ is that Bouda and Djama's $\dot{x}$ is not a group velocity while 
mine is. Hence, different quantum phenomena are predicted by the two representations.

This first explanation is also supported by Bertoldi, Faraggi and Matone.\ [\ref{bib:bfm}] 
\ \ They noted that, as the QEP is implemented in a Hamilton-Jacobi representation and 
not at a level of the equations of motion, the QSHJE must be solved before time 
parameterization may be introduced.  Time parameterization, consistent with QEP, is 
generated by Jacobi's theorem.\ [\ref{bib:bfm}] \ \ Bouda and Djama's Lagrangian 
formulation is effectively an attempt to introduce QEP at a level of the equations of 
motion. 

Second, underlying  Bouda and Djama's equation of motion, Eq.\ (B\&D's 17), is 
Faraggi and Matone's quantum transform [\ref{bib:fm}] 

\begin{equation}
 \hat{x}=\int ^x \frac{\partial S_0/\partial q}{\{2m[E-V(q)]\}^{1/2}}\, dq = \int^{S_0} 
\frac{ds}{[2m(E-V)]^{1/2}},
\label{eq:fm}
\end{equation}

\noindent which shows that the QSHJE admits a classical representation. Based on the 
QEP, Faraggi and Matone noted that the QSHJE is well-defined if and only if the ratio of 
the independent real solutions of the associated stationary Schr\"{o}dinger equation, 
$\phi _2(x)/\phi _1(x)$, is a local self-homeomorphism of the extended (includes $\pm 
\infty$) real line. We note that latent ratio, $\hat{\phi }_2(\hat{x})/\hat{\phi 
}_1(\hat{x})$, is not defined over the extended real line for the classical mechanics 
because classical trajectories may have turning points at finite values of $x$ and because 
the classical stationary Hamilton-Jacobi equation is a first-order differential equation.  In 
presenting the quantum transform, Faraggi and Matone noted that the integrand in Eq.\ 
(\ref{eq:fm}) becomes purely imaginary in the classical forbidden region and warn that 
classically forbidden regions correspond to critical regions for quantum coordinates.  The 
quantum transform is not a homeomorphism between $x$ and $\hat{x}$ over the 
extended real line.  This is another manifestation that QEP is incompatible with classical 
mechanics.  Lest we forget, classical mechanics innately cannot be consistent with the 
QEP.  Nevertheless, Bouda and Djama explicitly identified the quantum transform, Eq.\ 
(\ref{eq:fm}), as a coordinate transform and tacitly assumed it were consistent with the 
QEP by extending Eq.\ (B\&D's 17) into the classically forbidden region.\ [\ref{bib:bd}]\ 
\ Had the quantum transformation, Eq.\ (\ref{eq:fm}), been a QEP-preserving coordinate 
transformation, then classical mechanics would be consistent with the QEP because two 
different classical systems, ${\cal A}_{\mbox{\scriptsize classical}}$ and ${\cal 
B}_{\mbox{\scriptsize classical}}$, could be mapped into each other in four steps by 
QEP-preserving coordinate transformations.  First, the associated quantum systems ${\cal 
A}_{\mbox{\scriptsize quantum}}$ and ${\cal B}_{\mbox{\scriptsize quantum}}$ are 
already are consistent with QEP by coordinate transformations.  Second, the would-be 
QEP-preserving coordinate transformation, Eq.\ (\ref{eq:fm})  would relate ${\cal 
A}_{\mbox{\scriptsize quantum}}$ and ${\cal A}_{\mbox{\scriptsize classical}}$ 
consistent with QEP.   Third, the would-be QEP-preserving transformation, Eq.\ 
(\ref{eq:fm}) would likewise relate ${\cal B}_{\mbox{\scriptsize quantum}}$ and ${\cal 
B}_{\mbox{\scriptsize classical}}$ consistent with QEP.  And fourth, by the associative 
law, the relationship between the classical systems ${\cal A}_{\mbox{\scriptsize 
classical}}$ and ${\cal B}_{\mbox{\scriptsize classical}}$ would be consistent with 
QEP.  But this is a contradiction.  The resolution is that the second and third steps are 
invalid: Eq.\ (\ref{eq:fm}) is not a QEP-preserving coordinate transformation.  Bouda 
and Djama'a Eq.\ (8) corresponds to the quantum transform, Eq.\  (\ref{eq:fm}).

All discussions herein about phase and group velocities do not imply a pilot wave 
representation of quantum mechanics because analogous discussions are applicable to 
classical mechanics. 

I thank A.\ Bouda and T.\ Djama for their prompt and cordial council.  While we may 
still disagree on quantum trajectories, our discussions have been most fruitful in making 
our differences precise.   
 
\clearpage

\begin{enumerate}
\item \label{bib:bd} A.\ Bouda and T.\ Djama, Phys.\ Lett.\ {\bf A\ 285} (2001.) 27, 
quant-ph/0103071.
\item \label{bib:fm} A.\ Faraggi and M.\ Matone, Int.\ J.\ Mod.\ Phys.\ {\bf A 15} (2000) 
1869,  hep-th/9809127. 
\item \label{bib:c} R.\ Carroll, {\it Quantum Theory, Deformation and Integrability} 
(Elsevier, 2000, Amsterdam) p.\ 54.
 \item \label{bib:c1} R.\ Carroll, Can.\ J.\ Phys.\ {\bf 77} (1999) 319, 
quant-ph/9903081.
\item \label{bib:g} H.\ Goldstein, {\it Classical Mechanics}, 2nd ed.\ (Addison-Wesley, 
1980, Reading, MA) pp.\ 482--487.
\item \label{bib:f94} E.\ R.\ Floyd, Phys.\ Essays {\bf 7} (1994) 135.  
\item \label{bib:f00a} E.\ R.\  Floyd, Found.\ Phys.\ Lett.\ {\bf 13} (2000) 235, 
quant-ph/9708007.
\item \label{bib:f82} E.\ R.\ Floyd, Phys.\ Rev.\ {\bf D 26} (1982) 1339.
\item \label{bib:f00} E.\ R.\ Floyd,  J.\ Mod.\ Phys.\ {\bf A 15} (2000) 1363, 
quant-ph/9909072.
\item \label{bib:bfm} G.\ Bertoldi, A.\ E.\ Faraggi, and M.\ Matone, Clas.\ Quantum 
Grav.\ {\bf 17} (2000) 3965, hep-th/9909201.
\item \label{bib:lanczos} C.\ Lanczos, {\it The Variational Principles of Mechanics}, 4th 
ed. (Dover Publications, 1970, New York) pp.\ 161--164.
\item \label{bib:park} D. Park, {\it Classical Dynamics and Its Quantum Analogues}, 2nd 
ed. (Springer-Verlag, 1990, New York) p.\ 13. 
\end{enumerate}

\end{document}